\documentclass[twocolumn,showpacs,preprintnumbers,amsmath,amssymb]{revtex4}
\usepackage{graphics}% Include figure files
\usepackage{epsfig}
\usepackage{bm}% bold math
\begin{document}
%\preprint{Krenke01}
\title{Magnetic superelasticity and inverse magnetocaloric effect in Ni-Mn-In}
\author{Thorsten Krenke, Ey\"{u}p Duman, Mehmet Acet, and Eberhard F. Wassermann}
\affiliation{Experimentalphysik, Universit\"{a}t Duisburg-Essen,
D-47048 Duisburg, Germany}
\author{Xavier Moya, Llu\'{i}s Ma\~nosa and Antoni Planes}
\affiliation{Facultat de F\'isica, Departament d´Estructura i
Constituents de la Mat\`eria, Universitat de Barcelona, Diagonal
647, E-08028 Barcelona, Catalonia, Spain}
\author{Emmanuelle Suard and Bachir Ouladdiaf}
\affiliation{Institut Laue-Langevin BP 156, 38042 Grenoble Cedex
9, France}
\date{\today}% It is always \today, today,
       % but any date may be explicitly specified
\begin{abstract}
Applying a magnetic field to a ferromagnetic
Ni$_{50}$Mn$_{34}$In$_{16}$ alloy in the martensitic state induces
a structural phase transition to the austenitic state. This is
accompanied by a strain which recovers on removing the magnetic
field giving the system a magnetically superelastic character. A
further property of this alloy is that it also shows the inverse
magnetocaloric effect. The magnetic superelasticity and the
inverse magnetocaloric effect in Ni-Mn-In and their association
with the first order structural transition is studied by
magnetization, strain, and neutron diffraction studies under
magnetic field.
\end{abstract}


\maketitle

\section{Introduction}
Shape memory alloys exhibit unique thermomechanical properties
which originate from a martensitic transition occurring between
the austenite state with high crystallographic symmetry and a
lower symmetry martensite state \cite{Otsuka98}. These materials
are superelastic and can remember their original shape after
severe deformation. Superelasticity is related to the stress
induced reversible structural transition.

Research on shape memory alloys received significant stimulus
after the discovery of the magnetic shape memory (MSM) effect in
Ni$_2$MnGa \cite{Ullako96}. This effect arises from a magnetic
field induced reorientation of twin-related martensitic variants
and relies on high magnetocrystalline anisotropy. The driving
force is provided by the difference between the Zeeman energies of
neighboring martensite variants \cite{Ohand98,James98}. Giant
strains up to 10\% have been reported for off stoichiometric
Ni-Mn-Ga single variant crystals with the 14M modulated
martensitic structure \cite{Sozinov02}. Over the past decade, vast
amount of knowledge accumulated on the properties of Ni-Mn-Ga
Heusler alloys has enabled to foresee the possibility of employing
these alloys in device applications \cite{Kakeshita02}.

In applied magnetic fields, the martensitic start temperature
$M_s$ of Ni-Mn-Ga shifts to higher temperatures along with the
other characteristic temperatures martensite finish $M_f$,
austenite start $A_s$, and austenite finish $A_f$ \cite{Bozhko99}.
With this feature, it is possible to induce a reversible
structural phase transformation, whereby strain can be fully
recovered on removing the field without the necessity of
prestraining the specimen \cite{Dikshtein00}. In such magnetic
field induced superelasticity, the maximum field induced strain
relies on the difference in the crystallographic dimensions in the
martensitic and austenitic states. When a field of sufficient
strength is applied at a temperature corresponding initially to
the austenitic state, the shift in all characteristic temperatures
(therefore the shift in the hysteresis associated with the
transformation) can be large enough so that the martensitic state
is stabilized. However, experiments performed in fields up to 10 T
have shown that in the case of Ni$_{54}$Mn$_{21}$Ga$_{25}$ the
rate of shift is only about $\sim$1 KT$^{-1}$ \cite{Dikshtein00}.
Neutron diffraction experiments under magnetic field on an alloy
with similar composition confirm these results \cite{Inoue00}.

Parallel to the development of the understanding of the MSM effect
in Ni-Mn-Ga and exploiting giant strains for applications, search
for other MSM material also took up considerable place in the
research agenda \cite{Wuttig00}. In Ni-Co-Mn-In, it has recently
been reported that when a magnetic field is applied to a
pre-strained single crystal specimen, the strain is recovered with
a value that is nearly equal to the size of the pre-strain
\cite{Kainuma06}. Although this is a considerable step in the
search for magnetic superelasticity, a system in which
considerable length change occurs reversibly by applying and
removing a magnetic field without requiring pre-strain is still to
be found.

Recently, we have investigated a number of Ni-Mn based Heusler
systems other than Ni-Mn-Ga with the aim of finding ferromagnetic
alloys that undergo martensitic transformations and understanding
their properties around the transformation point
\cite{Krenke05a,Krenke06}. In Ni-Mn-Sn \cite{Krenke05b}, we have
come across an inverse magnetocaloric effect at temperatures in
the range of the first order martensitic transition with size
comparable to that of the archetype Gd$_5$(Si$_{1-x}$Ge$_x$)$_4$
system, which exhibits the conventional giant magnetocaloric
effect \cite{Pech97}.

Here, we demonstrate the presence of both magnetic superelasticity
and the inverse magnetocaloric effect in Ni-Mn-In in the range of
the martensitic transition. Large field induced strains in
polycrystalline Ni-Mn-In of magnitude similar to that in
polycrystalline Ni-Mn-Ga are found. We show in Ni-Mn-In that
instead of the large field induced strain being due to twin
boundary motion in the martensitic phase, it relies essentially on
the reverse field induced martensite-to-austenite transition.
Below, we present results of field dependent magnetization,
calorimetry, neutron diffraction, strain, and length change
measurements in magnetic field on a Ni-Mn-In alloy and discuss the
field induced strain and the inverse magnetocaloric effect in
relation to the field induced martensite-to-austenite transition.

\section{Experimental}
Arc melted samples were annealed at 1073 K under argon atmosphere
for two hours and quenched in ice water. Magnetization
measurements were carried out using a superconducting quantum
interference device magnetometer, and calorimetric measurements in
magnetic field were performed using a high sensitivity
differential scanning calorimeter \cite{Marcos03}. Neutron
diffraction in magnetic fields up to 5 T was performed on the D2B
powder diffractometer at ILL, Grenoble. The strain measurements
were made using conventional strain-gage technique in magnetic
fields up to 5T.

\section{Results}
\subsection {Calorimetry and magnetization}

\begin{figure}
\includegraphics[width=8cm]{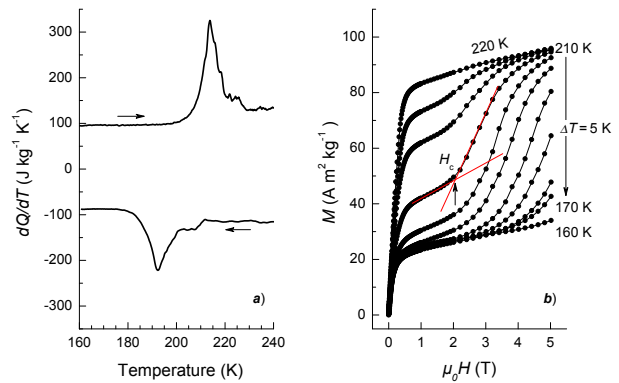}
\caption{\label{calorimetry} Features in the martensitic
transformation associated with temperature and magnetic field
dependence. a) Calorimetric curves across the martensitic
transition. The horizontal arrows indicate the direction of
temperature change. b) Magnetization as a function of magnetic
field measured on increasing field in the vicinity of the
martensitic transition. The red lines drawn through the data
points (shown only for the 200 K data) cross at a point
corresponding to the characteristic field around which the
metamagnetic transition begins to occur.}
\end{figure}

Ni$_{50}$Mn$_{50-x}$In$_{x}$ alloys undergo martensitic
transitions for about $x<16$ \cite{Sutou04,Krenke06}. Here we
concentrate on the magneto-structural coupling in the
ferromagnetic Ni$_{50.3}$Mn$_{33.8}$In$_{15.9}$ alloy, which has a
Curie point $T_C=305$ K and transforms martensitically on cooling
at $M_s=210$ K. The other characteristic temperatures defining the
temperature limits of the transition are $M_f=175$ K, $A_s=200$ K,
and $A_f=230$ K. These temperatures are determined from the
calorimetry data in Fig. \ref{calorimetry}a.

\begin{figure}
\includegraphics[width=8cm]{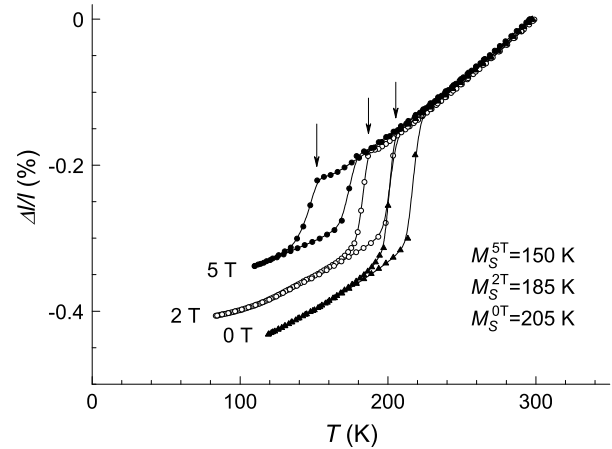}
\caption{\label{deltaL} The relative length change in constant
applied magnetic fields of 0 T, 2 T and 5 T. Arrows indicate the
positions of $M_s$.}
\end{figure}

In order to search for the presence of a coupling of the structure
with the magnetic degree of freedom within the temperature range
of the martensitic transition, we have studied the field
dependence of the magnetization $M(H)$. The data shown in Fig.
\ref{calorimetry}b are obtained in increasing field and decreasing
temperature. Here, the magnetizations in the temperature interval
$160\leq T\leq 210$ K initially show a tendency to saturate, but,
then, exhibit metamagnetic transitions in higher fields. The
characteristic field $H_c$ defining the transition point is
determined as the crossing point of the linear portions of the
curves. The transition shifts to higher fields with decreasing
temperature, and on removing the field, the magnetization returns
to its original value (see Fig. \ref{MH}). As will be shown with
neutron diffraction in external field, the metamagnetic transition
is associated with the onset of a field induced reverse
martensitic transition.

In Fig. \ref{deltaL}, we show the relative length change $\Delta
l/l$ as a function of temperature in different constant applied
magnetic fields. As the field increases, $M_s$ (indicated by
arrows) decreases by an amount of about $-10$ KT$^{-1}$. The other
characteristic temperatures are positioned in the conventional
manner around the temperature hysteresis loop, and all shift by
nearly the same amount in a given field. With increasing measuring
field, the difference in $\Delta l/l$ between the austenitic and
martensitic states decreases. The cause of this decrease is
associated with the crystallographic orientation of the easy axis
of magnetization within the orthorhombic structure of the
martensitic phase. This property is discussed separately in
reference \cite{Krenke07}.

\begin{figure}
\includegraphics[width=8cm]{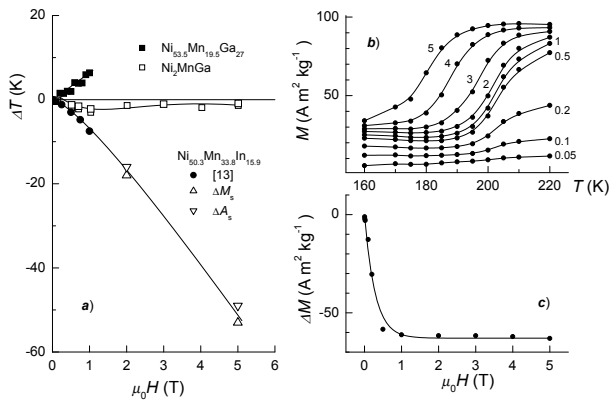}
\caption{\label{dT} Characteristic field dependent properties
around the martensitic transition. a) Shift in the martensitic
transition temperature as a function of magnetic field, for
Ni$_{50.3}$Mn$_{33.8}$In$_{15.9}$, Ni$_2$MnGa, and
Ni$_{53.5}$Mn$_{19.5}$Ga$_{27}$. b) Temperature dependence of the
magnetization obtained at selected fields from Fig.
\ref{calorimetry}b. The numbers refer to magnetic field values in
Tesla. c) The difference between the magnetizations in the cubic
and martensite phases. $\Delta M$ saturates at about 0.5 T. The
lines drawn through the data are guides.}
\end{figure}

In Fig. \ref{dT}a, we compare the magnitude of the shift of the
transition temperatures represented by $\Delta T$ as a function of
the external magnetic field $\mu_0H$ of the present Ni-Mn-In alloy
with those reported for Ni$_2$MnGa and
Ni$_{53.5}$Mn$_{19.5}$Ga$_{27}$; the latter exhibiting the
strongest field dependent transition temperature
\cite{Kim06,Jeong03}. Since the applied field shifts all
characteristic temperatures by the same amount, $\Delta T$ is the
change in any one of the characteristic transition temperatures in
applied magnetic field with respect to zero field. The shifts in
$M_s$ and $A_s$ of Ni-Mn-In determined from Fig. \ref{deltaL} are
shown with up and down triangles respectively. We have also
included data for $\mu_0H\leq 1$ T obtained from calorimetric
measurements under constant magnetic field \cite{Krenke06}.

Two significant features show up from Fig. \ref{dT}a: (i) The rate
of change of the transition temperature in
Ni$_{50.3}$Mn$_{33.8}$In$_{15.9}$ ($-10$ KT$^{-1}$) is higher than
in Ni$_{53.5}$Mn$_{19.5}$Ga$_{27}$ (6 KT$^{-1}$), and (ii) in
Ni-Mn-In, $\Delta T$ is negative, i.e., the transition temperature
decreases with increasing field. This is consistent with the fact
that the magnetization in the high temperature cubic phase is
larger than the magnetization in the martensitic phase as seen in
Fig. \ref{dT}b, where the temperature dependence of the
magnetizations at constant fields obtained from Fig.
\ref{calorimetry}b at selected fields are plotted. The difference
in the magnetization of the martensitic and austenitic states
around the transition temperature $\Delta M$ is plotted as a
function of applied field in Fig. \ref{dT}c.

\begin{figure}
\includegraphics[width=7cm]{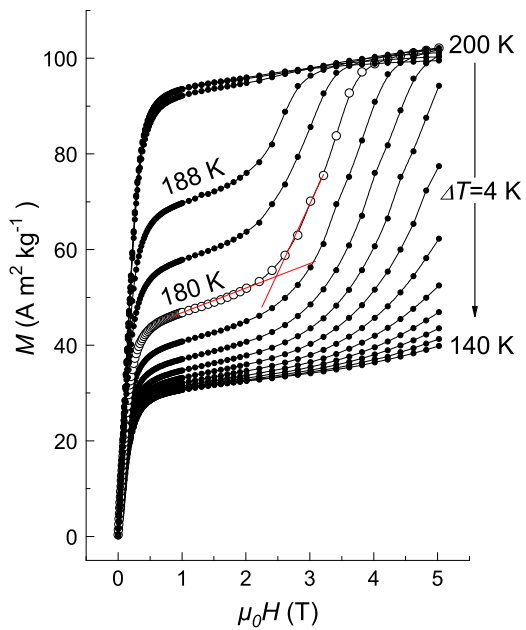}
\caption{\label{MHforNeut} The magnetic field dependence of the
magnetization for the sample used in the neutron diffraction
experiments. The crossing point of the red lines determine $H_c$.}
\end{figure}

\begin{figure}
\includegraphics[width=8cm]{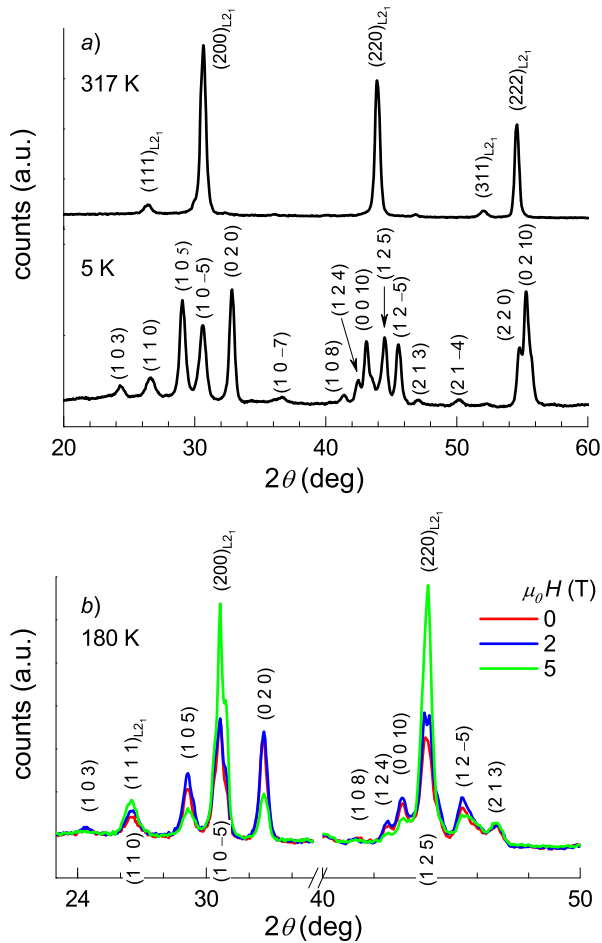}
\caption{\label{neut} Neutron diffractograms. a) Patterns at 317 K
(L2$_1$) and 5 K (10M martensite). b) The field dependence of the
diffraction pattern at 180 K showing the field induced
transformation from the martensite to the austenite state.}
\end{figure}

\subsection {Neutron diffraction in magnetic field}

To understand the properties of the transition observed in $M(T)$
and $M(H)$, we have undertaken powder neutron diffraction
experiments as a function of temperature and magnetic field. The
Ni$_{49.7}$Mn$_{34.3}$In$_{16.0}$ sample used for these
experiments has a composition that differs slightly from that used
in the measurements presented above and, therefore, the transition
temperatures are slightly shifted. Therefore, we plot in Fig.
\ref{MHforNeut} the $M(H)$ isotherms and will compare these data
to the neutron diffraction data.

Fig. \ref{neut}a shows neutron diffraction patterns at 5 K and 317
K. The pattern at 317 K generates from an L2$_1$ structure with a
lattice constant $a=0.6011$ nm. At 5 K, the pattern is that of a
10M modulated martensite structure having a monoclinic unit cell
with $\beta=86.97°$ and lattice constants $a=0.4398$ nm,
$b=0.5635$ nm, and $c=2.1720$ nm. Other than this slight
monoclinicity, the structure is orthorhombic having a shuffle
periodicity of 10 lattice planes in the [110] direction. In the
pattern at 317 K, some additional weak reflections can be
identified around 36° and 47°. These lie close to the positions of
the (1 0 $-$7) and (2 1 3) reflections of the martensitic phase at
5 K, but at slightly smaller angles due to thermal expansion, and
their presence is attributed to small amounts of mechanically
induced martensite formed on grinding the ingot for powder
specimen preparation.

At 180 K, which is a temperature that falls well in the range of
the transformation (Fig. \ref{MHforNeut}), we have studied the
evolution of the diffraction spectrum with applied magnetic field.
Fig. \ref{neut}b shows the spectrum in $2\theta$ ranges that
encompass the vicinity of the positions of the (200) and (220)
reflections of the L2$_1$ phase. As the magnetic field increases,
the intensities of these reflections grow at the expense of the
intensities of the (1 0 $-$5) and (1 2 5) reflections, which lie
nearly at the same positions as the (200) and the (220)
reflections of the L2$_1$ phase respectively. This shows that a
progressive magnetic field induced structural transition from the
martensitic to the austenitic state is taking place with
increasing magnetic field. In cases where the positions of the
reflections pertain only to the martensitic phase, e.g. (1 0 5),
(1 2 $-$5), etc., the intensity first increases with increasing
magnetic field up to 2 T and, then, decreases. The initial
increase is related to the increase in the magnetization in the
martensitic state at 180 K up to around $\mu_0H_c\approx 2$ T
(Fig. \ref{MHforNeut}). The subsequent decrease is associated with
the gradual decrease in the amount of martensite and the
stabilization of the L2$_1$ phase. However at 5 T, the reflections
associated with the martensitic phase do not disappear, although
their intensities are reduced. This indicates that the transition
is not complete in this field, and larger magnetic fields are
required to fully restore the L2$_1$ state at 180 K. The neutron
diffraction data indicate clearly a magnetic field induced reverse
transition and give evidence that the observed metamagnetic
transition in $M(H)$ is related to the reverse martensitic
transition.

\begin{figure}
\includegraphics[width=8cm]{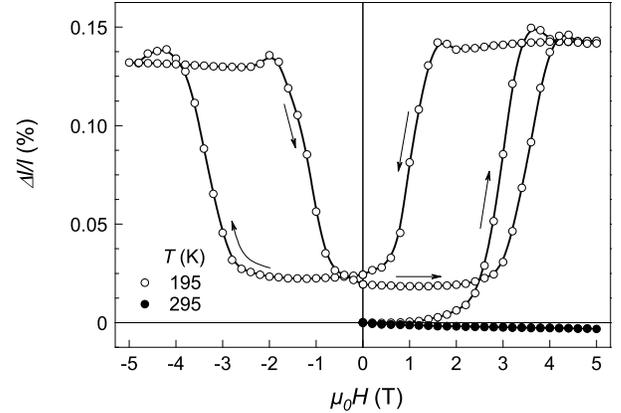}
\caption{\label{dl} Magnetic field dependence of strain at 195 K
($T<M_s$) and 295 K (L2$_1$). The strain recovers on removing the
field indicating magnetically superelastic behavior. Arrows show
the direction of field change.}
\end{figure}

\subsection {Magnetic field induced strain}

The magnetic field induced structural phase transition in the
present alloy can have important consequences on macroscopic
strains occurring during the application of the field. Fig.
\ref{dl} shows the results of magnetic field dependent strain
measurements, where the strain is defined as $\Delta
l/l=[l(H)-l_0]/l_0$. Here, $l_0$ is the length in the absence of
field and $l(H)$ the length in field. The sample is well within
the austenitic temperature range at 295 K (filled circles) and is
within the structural transition region at 195 K (open circles).
The small field induced strain at 295 K increasing negatively with
increasing field corresponds to the intrinsic magnetostriction of
the austenite, while at 195 K, a strain of about 0.14\% is reached
in the initial curve. After the first field cycle is completed,
the strain reduces to about 0.12 \% and remains constant at this
value, which is roughly the same as that attained in
polycrystalline Ni-Mn-Ga. However, the effect here is due to the
crystallographic transformation from martensite to austenite with
increasing field instead of a field induced twin boundary motion
that occurs within the martensitic state as in Ni-Mn-Ga. Although
hysteresis is observed in Fig. 6, the sample returns to its
zero-field length on removing the field.

\begin{figure}
\includegraphics[width=8cm]{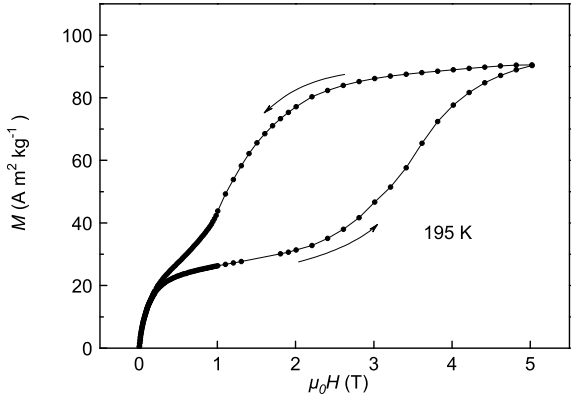}
\caption{\label{MH}The magnetic field dependence of the
magnetization at 195 K.}
\end{figure}

The features in the field dependence of the strain is reflected in
the field dependence of the magnetization at the same temperature
as seen in Fig. \ref{MH}. The $M(H)$ curves show metamagnetic
transitions around 2 T and 1 T for the increasing and decreasing
field branches respectively. These points correspond to the fields
where $\Delta l/l$ also changes rapidly. As in the case of $\Delta
l/l$, $M(H)$ also shows essentially no remanence and recovers its
zero-field value.

\begin{figure}
\includegraphics[width=8cm]{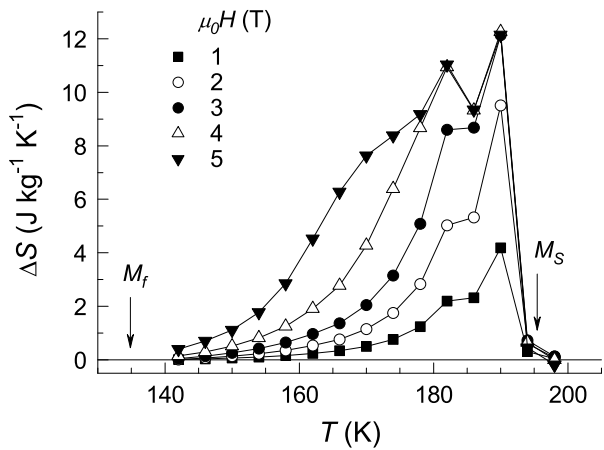}
\caption{\label{dS}Temperature dependence of $\Delta S$.}
\end{figure}

\subsection {Magnetocaloric effect}
Due to the first order magnetic field induced transition and
considerable difference in the magnetization of the martensitic
and L2$_1$ states at the transition temperature, substantial
magnetocaloric effects can be expected. The field induced entropy
change $\Delta S$ around the martensitic transition temperature
can be estimated from magnetization measurements by employing the
Maxwell equation \cite{Tegus02}

\begin{equation}
\Delta S(T,H) = \mu_0\int^{H}_{0}\left(\frac{\partial M}{\partial
T}\right)_{H} dH,
\end{equation}
from which the magnetocaloric effect can be evaluated by numerical
integration using the data in Fig. \ref{calorimetry}b. The
resulting $\Delta S$ in the temperature range 170 K $\leq T\leq
225$ K is plotted in Fig. \ref{dS}. The sign of $\Delta S$ is
positive for all temperatures indicating that an inverse
magnetocaloric effect is present, i.e. the sample cools on
applying a magnetic field adiabatically as in Ni-Mn-Sn
\cite{Krenke05b}. The maximum value of $\Delta S=12$
JK$^{-1}$kg$^{-1}$ is reached in a magnetic field of 4 T at about
205 K. Since there is no substantial change in the magnetization
above 4 T at this temperature, increasing the magnetic field any
further does not lead to any further increase in $\Delta S$. As
expected for magnetostructural transitions \cite{Casanova02}, this
value is larger than the transition entropy change \cite{Krenke06}
since it also includes the effect of magnetization changes with
temperature beyond the transition region.

\section{Discussion}

The origins of both distinct properties of the studied alloy,
namely, the field induced martensite to austenite transition and
the inverse magnetocaloric effect, are related to the lower value
of the magnetization in the martensitic phase with respect to that
in the austenitic phase. The difference in the magnetization can
be ascribed to the fact that in Mn based Heusler alloys, the
magnetic moments are localized mainly on the Mn atoms and the
exchange interaction strongly depends on the Mn-Mn distance.
Hence, any change in the distance caused by a change in the
crystallographic configuration can modify the strength of the
interactions leading to different magnetic exchange in each of the
phases. Indeed, it has been shown that in the case of a
Ni$_{0.50}$Mn$_{0.36}$Sn$_{0.14}$ alloy that short range
antiferromagnetism is present between Mn atoms located at the 4(b)
positions of the austenite phase which is then strengthened in the
martensitic state \cite{Brown06}. The present Ni-Mn-In alloy
transforms to the same martensitic structure as in
Ni$_{0.50}$Mn$_{0.36}$Sn$_{0.14}$. Therefore, the strong reduction
of the magnetization below $M_s$ in Ni-Mn-In can be expected to be
due to a similar cause. The presence of short range
antiferromagnetism in the ferromagnetic matrix leads to
frustration in the temperature range of the martensitic
transition. In such a frustrated system, the application of a
magnetic field can lead to the degeneracy of the spin states
giving rise to an increase in the configurational entropy that is
required for the observed positive $\Delta S$. However, the
quantitative details of the frustrated state and the microscopic
process leading to a positive $\Delta S$ remain to be described.

In Ni-Mn-Ga, giant strains are due to the reorientation of
twin-related variants in the martensitic state and recovery of
this strain is, in general, not achieved by simply removing the
field. By contrast, the magnetic superelastic effect and the
associated strain reported here for Ni-Mn-In is related to the
field induced structural transition. This enables to reversibly
induce and recover the strain by simply applying and removing the
field.

\section{Conclusion}

Ni$_{0.50}$Mn$_{0.34}$In$_{0.16}$ exhibits a magnetic field
induced structural transition from the martensitic state to the
austenitic state. The transition is directly evidenced by neutron
diffraction measurements under magnetic field. Here, other than in
Ni-Mn-Ga alloys, where the magnetization of the martensitic state
is higher than that in the cubic phase, the austenite is
stabilized by the application of a magnetic field. The shift of
the transition temperatures was found to be large and negative
with values up to about -50 K in 5 Tesla. Due to the reversible
magnetic field induced transition, magnetic superelasticity with
0.12\% strains occur. Other than in magnetic shape memory alloys,
where strain is mainly created by twin boundary motion, strain in
Ni$_{0.50}$Mn$_{0.34}$In$_{0.16}$ is caused by changes in lattice
parameters during the transition. Additionally, an inverse
magnetocaloric effect with a maximum value in the entropy change
of about 12 Jkg$^{-1}$K$^{-1}$ at 190 K and a minimum entropy
change of 8 Jkg$^{-1}$K$^{-1}$ in a broad temperature range $170 K
\leq T\leq 190 K$ is also found in this alloy.

\begin{acknowledgments}
We thank Peter Hinkel and Sabine Schremmer for technical support.
This work was supported by Deutsche Forschungsgemeinschaft (GK277)
and CICyT (Spain), project MAT2004-1291. XM acknowledges support
from DGICyT (Spain).
\end{acknowledgments}

\end{document}